Review Article for *Proceedings of the Institution of Mechanical Engineers, Part A: Journal of Power and Energy:*

Converting energy from fusion into useful forms

M. Kovari*, C. Harrington, I. Jenkins, C. Kiely

EURATOM/CCFE Fusion Association, Culham Science Centre, Abingdon, Oxon, OX14 3DB, UK

*Corresponding author.  Tel.: +44 (0)1235-46-6427.  E-mail address: michael.kovari@ccfe.ac.uk

**Abstract**

If fusion power reactors are to be feasible, it will still be necessary to convert the energy of the nuclear reaction into usable form.  The heat produced will be removed from the reactor core by a primary coolant, which might be water, helium, molten lithium-lead, molten lithium-containing salt, or $CO_2$.  The heat could then be transferred to a conventional Rankine cycle or Brayton (gas turbine) cycle.  Alternatively it could be used for thermochemical processes such as producing hydrogen or other transport fuels.   Fusion presents new problems because of the high energy neutrons released. These affect the selection of materials and the operating temperature, ultimately determining the choice of coolant and working cycle. The limited temperature ranges allowed by present day irradiated structural materials, combined with the large internal power demand of the plant, will limit the overall thermal efficiency. The operating conditions of the fusion power source, the materials, coolant, and energy conversion system will all need to be closely integrated.

Keywords: fusion power plant, tritium, tokamak, Rankine cycle, Brayton cycle, helium, $CO_2$, lithium-lead, efficiency.

# 1. Introduction

Interest in the possibility of controlled nuclear fusion of hydrogen isotopes dates back to before the end of World War II. The one reaction that is nearest to commercially relevant achievement is the fusion of deuterium and tritium to make helium. Many proposals have been put forward, but three reactor concepts have come to dominate research work: the tokamak, the stellarator, and laser inertial fusion. In a tokamak a very low density mixture of deuterium and tritium in the form of a plasma at about $2 \times 10^8$ Kelvin is confined in a toroidal shape by a magnetic field. A current running around the torus generates an additional magnetic field, so the field lines wind around the machine in a helical fashion. An alternative design, the stellarator, is similar in principle, but there is no electric current in the plasma, giving greater stability. Instead the plasma itself is formed into a twisted shape, requiring complex non-axisymmetric magnets. In laser fusion, a pellet of deuterium and tritium is compressed and heated by a set of lasers.

The tritium fuel, which does not occur naturally in significant quantity, must be manufactured by allowing the neutrons to be absorbed by lithium, which reacts to form tritium. This process is known as breeding. A final requirement is to extract the energy released in the fusion reaction and convert it to useful power. In the deuterium/tritium reaction 80% of the energy appears as the kinetic energy of the neutrons released. The neutrons must be slowed down, converting their energy into heat. The remaining 20% appears as heat incident on the inner surfaces of the reactor.

Neutrons from deuterium-tritium fusion are born with a high energy (14 MeV). They are highly damaging to materials, both through atomic displacement and by transmuting the elements into less desirable ones. High temperatures can help to anneal radiation damage, restoring ductility in some materials. There are therefore close links between the choice of materials, coolant and thermodynamic cycles.

The recent European Fusion Roadmap[1] suggests that with sufficient funding and favourable experimental results, fusion electricity production could be demonstrated at the end of 2048. (ITER, the tokamak under construction in France, will be an experimental machine only. It is not intended to demonstrate the thermodynamic cycle so its cooling systems dump the heat into the environment through cooling towers.) To roll fusion power out worldwide would then require the manufacture of sufficient tritium, either by extra breeding in the first reactors, or using fission reactors.

## 2.     A fusion power plant

We can now outline the components of a fusion power plant (Figure 1 and 2). In magnetic confinement fusion the hot fuel is surrounded by a breeder blanket containing lithium and a neutron multiplier such as beryllium or lead. A bonus is that the breeding and multiplying reactions also release energy, so that total energy production is about 120% of the energy released by fusion alone. The nuclear energy is deposited in the blanket as heat, and is removed by a coolant.

As in a Pressurised Water Reactor (PWR), both the reactor and the primary heat exchanger must be inside a biological shield, since the reactor is a powerful source of neutrons, and the coolant will contain activation products as well as tritium. The primary coolant has to collect heat from several distinct components, which receive different power densities and radiation levels. Listing from the plasma outwards, they are the "first wall" (the surface facing the plasma), the blanket, the radiation shield (which protects the vacuum vessel), and the vacuum vessel. In addition the ions leaking from the plasma, including both fuel and helium "ash", must be collected in a structure known as a divertor. In principle a direct Brayton cycle is possible with a single coolant (as has been proposed for high temperature fission reactors), but indirect cycles with primary and secondary coolants have received most attention, since the tritium concentration in the primary coolant will be high. Direct conversion of energy from non-thermal forms into electricity has been proposed using electromagnetic techniques, but is not currently favoured.

In inertial confinement the plant layout is similar, but with a spherical vacuum chamber into which the fuel pellets are injected. There are no coils, but the blanket, shielding and containment are still required. In this case the blanket must have many ports through which the laser beams pass.

Unlike a conventional fossil fuel or fission plant, the fusion plant may have an internal power demand that is a substantial fraction of the gross electricity production – either to drive the circulating current in a tokamak, or to power the lasers in an inertial fusion plant. (A stellarator does not have this issue.) This power is also required to start the reactor, so that "black start" (starting up the plant with no external power source) would be very challenging.

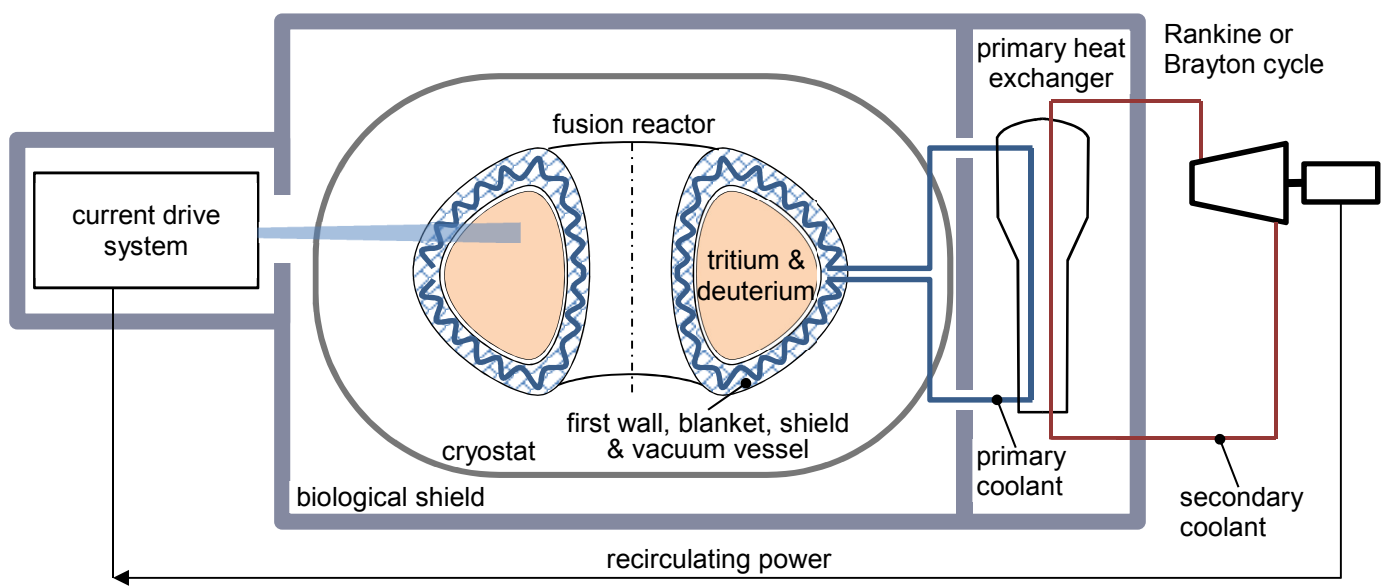

**Figure 1**. Schematic of a possible tokamak fusion power plant. The space within the cryostat is occupied by the superconducting coils (not shown). The divertor and the plant for extracting tritium from the blanket are not shown.

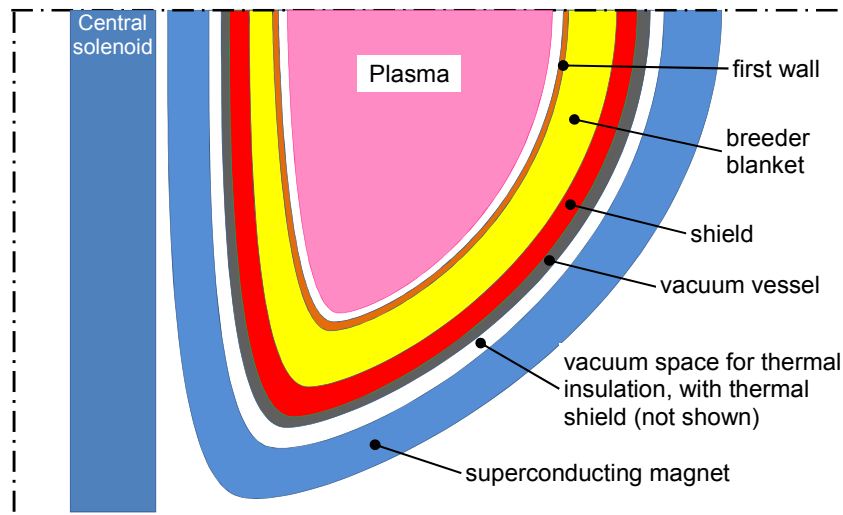

**Figure 2**. Detail of the lower right portion of the cross-section. The poloidal field coils are not shown. The machine axis is shown on the left.

## 3. The breeding blanket and primary coolants

The problem of cooling a fusion reactor is superficially similar to that of a fast breeder fission reactor[2]: the coolant must not absorb or slow down the neutrons too much, as that would reduce the breeding of tritium, and the materials must retain their integrity under intense fast-neutron irradiation. However, the use of a neutron multiplier means that light water can be used as a coolant, even though it absorbs and moderates neutrons, which is not possible in a fast breeder. Primary coolants considered to date include water, helium, molten lithium-lead, molten lithium-containing salt, or $CO_2$. The space available for the blanket is limited on the outboard side by the need to fit it inside the superconducting magnetic coils, and on the inboard side by the need to fit it inside the hole in the torus. The outboard blanket needs to be ~50-100 cm thick to breed as much tritium as is consumed. The coolant pressure will be limited by the need for very high reliability and by radiation damage to the pipes. The power density of neutron heating in the blanket is up to ~ 10 MW/m$^3$, but the first wall will in addition receive a direct heat load of the order of 0.5 MW/m$^2$.

There must be a system for removing tritium from the coolant in the primary circuit, and probably also in the secondary circuit, but this is not shown in the flow diagrams in this paper. The power required is not known, and is generally ignored.

The structural **materials** selected play a central role in determining the thermodynamic parameters[3]. Reduced activation ferritic–martensitic steels (RAFM) have a narrow temperature window: radiation damage causes embrittlement below 200-250ºC, while strength declines above 550ºC. Tungsten has an exceptionally high melting point and resistance to erosion by incoming ions, but the minimum temperature to avoid severe radiation hardening embrittlement is expected to be 900 ± 100°C. While this is an authoritative estimate, tungsten is still popular in paper design studies, which often assume acceptable behaviour at much lower temperature.

Table 1. Comparison of coolants[4, 5]

| | Pressure | Start temp | End temp | Enthalpy increase / volume | Mean viscosity | Mean thermal conductivity | Figures of Merit[6] | |
|---|---|---|---|---|---|---|---|---|
| | | | | | | | Heat transfer* | Pumping power** |
| | MPa | °C | °C | J/m$^3$ | 10$^{-6}$ Pa·s | W/m/K | Relative to water at 25°C, 0.1 MPa | |
| Water | 15 | 265 | 325 | 2.2E+08 | 90 | 0.57 | 2.78 | 2.8 |
| Steam | 8 | 300 | 500 | 1.5E+07 | 24 | 0.070 | 0.17 | 5.06E-04 |
| Helium | 8 | 300 | 500 | 5.1E+06 | 35 | 0.28 | 0.16 | 1.17E-04 |
| CO$_2$ | 8 | 300 | 500 | 1.3E+07 | 31 | 0.049 | 0.13 | 2.46E-04 |
| Flibe Li$_2$BeF$_4$ | | 500 | 700 | 9.2E+08 | 10200 | 1.0 | 0.51 | 0.50 |
| Pb-17Li | | 700 | 1100 | 6.0E+08 | 650 | 25 | n.a. | n.a. |

\* Related to the rate of heat transfer per unit pumping power for a given geometry.

\*\* Inversely related to the pumping power required to transport a given amount of energy.

**Water** is unsurpassed as a heat transfer medium (Table 1). Whether as steam or liquid it can transfer and carry more heat per unit pumping power than other coolants. Removing tritium from water poses a particular problem as the absorbed tritium has to be separated from a huge quantity of stable hydrogen, while isotopic exchange will ensure that the tritium does not remain chemically distinct. Water and especially steam are likely to attack the pipes chemically, and may well dissolve highly activated corrosion products. The oxygen forms $^{16}$N when irradiated by neutrons. This reaction is not very significant for fission reactors as it has a neutron energy threshold of 10.5 MeV, but becomes important for fusion, which produces neutrons of 14 MeV. The $^{16}$N has a half-life of only 7.1s, but it emits gammas at 6.1 and 7.1 MeV.

A conceptual study of a water-cooled fusion reactor[7] proposed the parameters shown in Table 2, similar to those of a PWR.

Table 2. Cooling system and Rankine cycle parameters for a water-cooled fusion reactor.

| **Primary circuit** | |
|---|---|
| Total nuclear heat output | 5300 MW |
| Heat deposited in first wall and blanket | 4300 MW |
| Water output temperature | 325 °C |
| Water input temperature | 265 °C |
| Water pressure | 15 MPa |
| Flow rate | 13000 kg/s |
| **Secondary circuit (see section 4)** | |
| Steam generator outlet temperature | 285 °C |
| Condensate inlet temperature | 230 °C |
| Operating pressure | 69 bar(a) |
| Flow rate | 6 x 402 kg/s |
| Gross output | 1800 MWe |
| Gross efficiency | 35% |
| Net output | 1550 MWe |
| Net efficiency = net electric / nuclear heat | 29% |

The divertor supports a larger surface heat flux than the rest of the first wall. In the design described in Table 2 the divertor consequently uses cooling pipes made from CuCrZr because of its high thermal conductivity, surrounded by tungsten armour. To maintain its strength the copper alloy is kept below 300ºC, so the divertor cooling water is cooler than the blanket water, and is used for preheating the feedwater in the working cycle (Figure 4).

**Helium** has excellent thermal conductivity, is chemically inert and does not become radioactive under neutron irradiation. In other ways its properties are poor – it is an expensive and limited resource, it is likely to escape from any small leaks, and its density is low and compressibility is high, so high pumping power is required. At least seven helium-cooled fission reactors have been built and operated, however, at up to 330 MWe, so the technology is well-established[8].

A possible reactor with helium as primary coolant [9] is outlined in Table 3 and Figure 3. In contrast to the water-cooled study above, here the helium that cools the divertor must enter and exit at a very high temperature because the divertor is made of tungsten – entering at 541ºC and exiting at 717ºC. In these conditions the tungsten remains within its operating window, assumed in this study to be 600–1300ºC. Even so the divertor is only expected to survive about 10–100 cycles between room and operating temperature.

The design allows the high temperature helium from the divertor to superheat the steam generated by the blanket loop. Energy deposited by neutrons in the outer parts of the radiation shield and the vacuum vessel itself counts as low-grade heat and is lost. The pumping power is extremely high (but of course most of this power is deposited in the coolant).

Table 3. Primary coolant parameters for a helium-cooled fusion reactor and an advanced reactor cooled by lithium-lead.

| Primary coolant | Helium [9] | Pb-17Li [10] |
|---|---|---|
| Total nuclear heat output (MWth) | 5000 | 2796 |
| Pressure (MPa) | 8 | 1.5 |
| Flow rate (total) (kg/s) | 5000 | 36900 |
| Coolant inlet and outlet temperatures in blanket (°C) | 300/500 | 700 /1100 |
| Pressure drop in the blanket (MPa) | 0.32 | 0.19 - 0.85 |
| Pumping power (MWe) | 400 | 12 |
| Power for current drive (MWe) | 430 | 101 |
| Net electric power (MWe) | 1500 | 1530 |
| Net efficiency = net electric / total nuclear heat | 30% | 55% |

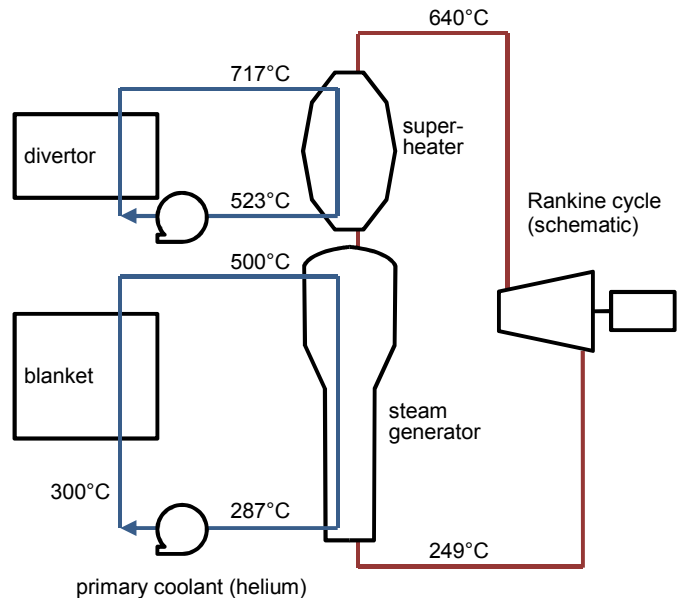

**Figure 3**. A helium-cooled reactor cycle. Note that the helium heats up significantly as it goes through the pumps.

**Lithium-lead** could combine the functions of tritium breeder and neutron multiplier. In principle it can even act as primary coolant ('"self-cooled blanket"), or it can be cooled by water or helium. In a compromise "dual-cooled" design, the lithium-lead would be circulated fast enough to allow removal of the heat deposited in it, but an additional helium circuit would cool the first wall and the structural components. Liquid metals have even been proposed for the first "wall", directly exposed to the plasma. Liquid metals have the advantages of high boiling point at low pressure, and high thermal conductivity. A major drawback is that they conduct electricity. The high magnetic fields of a tokamak or stellarator will induce eddy currents in moving metals, and these currents will, by Lenz's law, generate magnetic forces which act to impede the flow. Not only does this magnetic drag make it difficult to pump the metal around the circuit, but it also impairs heat transfer by suppressing turbulence. These effects can, however, be reduced by using insulating pipes, or insulating coatings or inserts in the pipes to prevent current flow from liquid to pipe.

Parameters for a lithium-lead-cooled reactor, based on technology well ahead of what is available today, are given in Table 3. The mass flow rate is very high because of the low specific heat capacity. The melting point of 234ºC for Pb-17Li is well below the operational temperature, but remains an inconvenience for starting up and shutting down. This study assumed that the structural material of the blanket would be a composite of silicon carbide reinforced with fibres of the same material, as used for body armour.

**Molten salts** are used industrially for heat transfer. Flibe, a mixture of lithium fluoride (LiF) and beryllium fluoride (BeF2), has been proposed as a fission reactor coolant. The 2:1 mixture with proportions $Li_2BeF_4$ has a melting point of 459°C, and a boiling point of 1430°C.[11] The eutectic mixture is slightly greater than 50% $BeF_2$ and has a lower melting point of 360°C. For fusion Flibe has the additional advantage that it contains both lithium and beryllium. It does not react with air or water, has low vapour pressure, and is chemically compatible with RAFM. Its viscosity and melting point are high, but it has been proposed as both breeder and coolant for a stellarator[12], perhaps combined with a helium Brayton cycle.

## 4. Secondary coolants and electricity generation

The **steam** Rankine cycle provides a comfortable place to rest before exploring more exotic options.

A fusion plant whose primary coolant can reach 500-700°C could use either a subcritical steam cycle, in which the high temperature is used to superheat the steam after boiling, or a supercritical cycle. In either case the high temperature allows good cycle efficiency, although the net efficiency is diminished by the pumping power for the helium primary coolant.

In contrast, if the primary coolant is limited to lower temperatures, such as the water-cooled concept described above, the boiling point achievable at realistic secondary water pressure would be close to the maximum temperature. As a result there is no potential for superheating the steam, which now enters the turbine at its saturation point. An example is shown in Table 2 and Figure 4. (The net plant output is based on a speculative estimate of the internal power demand.) The efficiency is naturally poor because of the low temperature.

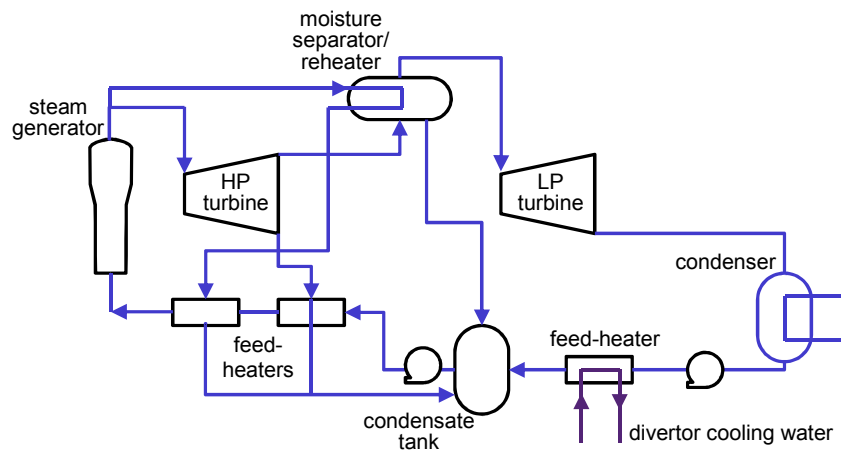

**Figure 4**. A Rankine cycle for a water-cooled reactor (simplified)[7]. The divertor, assumed here to be operating at relatively low temperature, is used to pre-heat the condensate.

Figure 4 shows the use of techniques for boosting the efficiency of the standard Rankine cycle, including reheating the steam after it has partially expanded, and feedheating, in which the water is heated before entering the steam generator using steam extracted from other points in the cycle. As in a PWR, the moisture-separator is needed to remove the water droplets condensing in the high-pressure turbine.

**Carbon dioxide** has a good track record as a primary coolant in fission plants. In the secondary circuit it has the potential for high efficiency due to the low compression work near the critical point (7.38 MPa, 31ºC). Consider for example a dual-cooled reactor in which lithium-lead is circulated so that it acts as a primary coolant as well as breeder, but helium is also used to cool the structural components[13]. As before we assume that the divertor is much hotter than the blanket (helium temperatures reaching 800ºC and 400ºC respectively, while the liquid metal reaches 700ºC). Taking account the relatively low inlet helium temperature (300ºC), a $CO_2$ recompression Brayton cycle with a Rankine bottoming cycle was proposed, giving 47% gross efficiency. Electrical power for current drive and for pumping the primary coolants still has to be subtracted from this, so it is hard to make a comparison with a simple Rankine cycle.

A **helium** Brayton cycle, with the same assumptions as for the $CO_2$ study above, was able to use the high grade heat efficiently, using a recuperator to cool the gas leaving the turbine, followed by an additional precooler, and two stages of compression (driven by the turbine) with intercooling. The lower grade helium from the blanket, however, could only be effectively used with a steam cycle. Even when the two cycles were combined the overall efficiency was significantly less than with $CO_2$.

## 5. The intermittent (pulsed) reactor

A fusion power plant based upon the tokamak principle requires a source of current in the plasma. Present experimental machines, such as the European tokamak "JET", use a solenoid through the centre of the machine to induce a current, the plasma acting as a very low resistance single turn secondary coil of a transformer. However, such a current can be induced in the plasma only for a limited time, depending on the size of the solenoid. Additional sources of current have been developed, using microwaves or injected beams of neutral atoms, but for these systems to provide sufficient current in steady state with sufficient power efficiency will take considerable development.

The tokamak is thus, at present, an inherently pulsed device, and both ITER and the proposed near-term European demonstration power plant are designed to operate as such with a dwell period in between the pulses allowing the recharge of the central solenoid. No other power generating technologies operate in an intentionally pulsed manner and this would certainly affect components such as steam turbines which are not generally designed for frequent cycling. Research, in partnership with industry, is presently under way to understand how to mitigate such risks and attempt to optimise efficiency in pulsed operation. With a sufficiently large central solenoid the pulse length could theoretically be as long as eight hours, requiring perhaps 15 minutes to recharge.

Another consideration is the profile of electricity output to the grid. The first one or two power plants are likely to be treated as special cases which will operate in close co-ordination with the electricity grid operator. They are likely to be of the order of 500 MWe and so not crucial to grid stability. Nevertheless, when a pulsed power plant is between fusion burns, the grid will need to compensate in some way, either using energy storage, or by rapidly ramping up spare capacity. In addition, power is required to restart the fusion reactor. While there are many types of energy storage available, storing heat using molten salt might be particularly suited to a thermal power plant such as a fusion plant, providing temperatures of the order of 500ºC can be achieved. Molten salt energy storage is a well-developed technology, with solar plants in Spain equipped with storage capacity of 50 MWe for 7.5 hours. The need to transfer heat between water, which undergoes a phase change, and molten salt, which does not, poses substantial difficulties for integration into the Rankine cycle. This pinch point problem is discussed in the solar power literature [14]. The problem is particularly acute if the steam both charges and discharges the salt. An alternative is for the salt to be heated directly by the primary coolant (perhaps helium), which eliminates one pinch point. Alternatively the energy could be used to manufacture products that can be stored in bulk, such as fresh water or hydrogen.

## 6. Hydrogen Production

Currently, 45 Mt of hydrogen is produced worldwide each year[15], mostly through steam-methane reforming[16], but this has a large $CO_2$ footprint.

Hydrogen production processes can be split into thermo-chemical processes and hybrid processes that utilise both heat and electricity. The sulphur-iodide thermo-chemical process and the copper-chloride hybrid process have the most promise, but maximum temperature requirements are 900˚C and 500˚C respectively[17, 18]. The Chinese FDS-III reactor concept is focusing on a lithium-lead coolant with a blanket temperature of 1000˚C in order to use the sulphur-iodide process [19]. However, there are enormous material challenges at this temperature and no operating power plants of any kind achieve temperatures of this magnitude. Although the copper-chloride cycle requires an input of electricity, it has a much more achievable top temperature which is comparable to Advanced Gas Reactors (AGRs) operating today.

## 7.     Desalination

Current worldwide water consumption is 5.68 trillion $m^3$ per year. Of this, only 29.2 billion $m^3$ is desalinated water. With worldwide population growth at 80 million a year, water consumption will certainly increase. Countries such as China are investing in desalination to meet this growing demand[20].

There are two dominant techniques for water desalination: reverse osmosis and Multi Stage Flash Distillation (MSFD)[21]. Reverse osmosis requires pumping water at a pressure of around 70 bar through a membrane which separates the salts from the freshwater. A water pre-treatment system is required and the membrane life is around 5-7 years[22]. MSFD is a thermal process that requires low-grade heat with a maximum temperature of 120˚C. The water is evaporated from the salts in many stages and collected at each stage.

MSFD requires a higher energy input than Reverse Osmosis but it is reliable and proven, does not require pre-treatment and can be coupled with power generation. For a fusion reactor (or other thermal power plant), MSFD could utilise the low-grade heat in the working fluid after it has passed through the turbines (with some reduction in electricity output).

## 8. Summary

Experiments on machines such as JET and simulations all suggest that the experimental reactor ITER may well demonstrate the liberation of fusion power at ~ 400 MWth, and neutronics models suggest that adequate tritium production could be feasible in a power plant. In a power plant the fusion energy could be converted to electricity using conventional methods, or it could be used for desalination or the production of transport fuels. The high energy neutrons, the high surface heat loads, and erosion due to energetic particles will have a strong influence on the materials, temperatures and pressures that can be used. Possible approaches are described here but there is as yet not a fully consistent solution for engineering design, coolants and working cycle.

**Acknowledgments**

This work was part funded by the RCUK Energy Programme [grant number EP/I501045] and the European Communities under the Contract of Association between EURATOM and CCFE. The views and opinions expressed herein do not necessarily reflect those of the European Commission.